\documentclass[twocolumn]{aastex7}
\usepackage{mathtools}

\begin{document}

\title{Revisiting Sub-Neptune Population Synthesis: The Role of Core Luminosity in Boil-off, Photoevaporation, and the Radius Distribution}

\correspondingauthor{Yao Tang}
\email{yaotang@ucsc.edu}

\author[0000-0003-3980-7808]{Yao Tang}
\affiliation{Department of Astronomy and Astrophysics, University of California, Santa Cruz \\
1156 High Street, Santa Cruz, CA 95064, USA}
\email{yaotang@ucsc.edu}

\author[0000-0002-9843-4354]{Jonathan J. Fortney}
\affiliation{Department of Astronomy and Astrophysics, University of California, Santa Cruz \\
1156 High Street, Santa Cruz, CA 95064, USA}
\email{jfortney@ucsc.edu}

\author[0000-0002-7030-9519]{Natalie Batalha}
\affiliation{Department of Astronomy and Astrophysics, University of California, Santa Cruz \\
1156 High Street, Santa Cruz, CA 95064, USA}
\email{nabatalh@ucsc.edu}

\author[0000-0001-5061-0462]{Ruth Murray-Clay}
\affiliation{Department of Astronomy and Astrophysics, University of California, Santa Cruz \\
1156 High Street, Santa Cruz, CA 95064, USA}
\email{rmc@ucsc.edu}

\begin{abstract}
The observed bimodal radius distribution of close-in small planets provides important constraints on their formation and evolution. Two atmospheric escape mechanisms, photoevaporation and core-powered mass loss, have been widely invoked to explain this distribution. In this work, we revisit the physical assumptions underlying these models and investigate the role of core luminosity in driving outflows and shaping the radius distribution. We develop a population synthesis framework based on the SEAMIST evolution model, which self-consistently couples the two mass-loss mechanisms while accounting for envelope metallicity. We find that envelope metallicity strongly affects the resulting radius distribution, with 50-100$\times$ solar metallicity providing a good match to the observed population. Boil-off generally dominates atmospheric mass loss, while photoevaporation typically removes at most 3 wt\% of the envelope mass. We find that core luminosity contributes only $\sim$10-20\% of the thermal energy available for envelope re-equilibration during boil-off, producing an absolute mass-loss enhancement of only $\sim$0.4wt\%. Its effect is larger during photoevaporation, where the longer mass-loss timescale allows core heating to accumulate. Nevertheless, core luminosity has little impact on the long-term planetary radius distribution. These results suggest that core-powered mass loss is better understood as a modification to atmospheric escape through its lower-boundary conditions rather than as a distinct mass-loss mechanism. In our best-fit model, the radiative atmosphere contributes a significant $\sim$20\% of the total optical radius. Consequently, we highlight the need for improved radiative-atmosphere modeling in sub-Neptune evolution models to better understand the observed population distribution.
 
\end{abstract}


\section{Introduction} \label{sec:intro}
Since the launch of \emph{Kepler}, thousands of transiting exoplanets have been discovered through the transit method. Among them, planets with radii $\lesssim4R_\oplus$ and orbital periods shorter than 100 days constitute one of the most abundant populations \citep{Borucki11,Batalha12}. An analysis of these small, highly irradiated planets by \citet{Fulton17} revealed a pronounced dichotomy in their radius distribution: the population is separated into a smaller-radius ($1-1.7R_\oplus$) population of predominantly high-density planets and a larger-radius population ($2-4R_\oplus$) of lower-density planets, with a deficit of planets at intermediate radii. Prior to this discovery, theoretical studies \citep{Owen13,Lopez13} had predicted such a dichotomy by proposing that initially low-density, gas-rich planets could undergo efficient atmospheric mass loss and completely lose their gaseous envelopes, leaving behind bare, rocky cores. The resulting transformation from a gas-rich planet to a terrestrial planet can produce a dramatic reduction in planetary radius. Owing to their characteristic resemblance to Neptune and Earth, and their typical masses of $\sim1-20M_\oplus$, these two populations have subsequently been referred to as \emph{sub-Neptunes} and \emph{super-Earths}, respectively.

This observed radius distribution and its possible evolutionary origin have since motivated extensive theoretical \citep{Lopez14,Jin14,Chen16,Owen17,Ginzburg18,Rogers20} and observational efforts \citep{vaneylen18,Hardegree19,Hsu19,Kunimoto20,Bryson20,Dattilo23,Gillis26} to understand the evolution of close-in small planets. In particular, three fundamental questions arise: How do planets form and evolve into the observed radius distribution? Which atmospheric mass-loss mechanisms dominate this evolution? And what are the underlying distributions of planetary properties, such as core mass, initial envelope mass, and envelope composition, that give rise to the observed population?

Two atmospheric mass-loss mechanisms, photoevaporation \citep{Lammer03,Yelle04,RMC09} and core-powered mass loss \citep{Ginzburg16}, have been widely discussed in the literature and have both been shown to reproduce a bimodal radius distribution resembling that observed \citep{Owen17,Ginzburg18,Gupta19}. More generally, atmospheric outflow requires two essential ingredients: a pressure gradient that drives the expansion of the atmosphere and a source of energy that sustains the outflow. An atmosphere can remain hydrostatically confined only when the external pressure at its outer boundary is sufficiently high to provide confinement against the pressure gradient within the atmosphere. As the external confining pressure decreases, or as atmospheric heating increases the pressure scale height, the atmosphere expands to maintain hydrostatic equilibrium. When the atmosphere becomes sufficiently extended that hydrostatic confinement can no longer be maintained, an outflow develops. However, expansion of gas against the planetary gravitational potential require energy. Without a sufficient energy supply, the gas cools and the outflow cannot be sustained. 

Photoevaporation is driven by the pressure gradient established by heating from extreme-ultraviolet and X-ray photons (collectively, XUV), which deposit energy in the upper atmosphere and can heat the gas to temperatures of a few thousand to more than $10^4$ K \citep{RMC09,Lammer16,Tang25b,Broome25,Kubyshkina26}. In contrast, core-powered mass loss, despite its name, is powered primarily by stellar bolometric irradiation. When a planet has a sufficiently extended atmosphere, bolometric irradiation can establish, through the resulting equilibrium temperature, a pressure gradient strong enough to drive a hydrodynamic outflow in the form of a Parker wind \citep{Parker1958}. In this sense, the outflow resembles the inverse of the gravitationally bound flow in Bondi accretion, with gas escaping from rather than accreting onto the planet. 

According to \citet{Ginzburg16}, core-powered mass loss affects the outflow through the lower boundary condition. First, core luminosity regulates the rate at which gas material is supplied to the outflow, heating the envelope and lifting material within it, thereby controlling how efficiently the envelope can re-equilibrate at its outer boundary, where it connects to the outflow. This process, termed \emph{envelope re-equilibration}, forms the basis of their mass-loss prescription. Moreover, \citet{Ginzburg16} showed that when the envelope is sufficiently small, the core can contain more energy than the envelope itself. In this regime, the cooling of the core can efficiently supply energy to the envelope, maintaining its thermal structure and enabling a sufficiently large pressure gradient to sustain the outflow for billions of years.

However, a critical reassessment of the envelope re-equilibration assumption by \citet{Tang24} (TFM24, hereafter) shows that the envelope redistributes its energy and mass through envelope convection that operates on a timescale much shorter than the mass-loss timescale. As long as the envelope remains adiabatic, this process does not require an external energy source. On this characteristic timescale, the lifted envelope material responds to the decreasing external pressure by expanding adiabatically: as the gas expands, its pressure and temperature decrease, and its internal thermal energy is converted into gravitational potential energy associated with lifting gas to larger radii. The entropy of the envelope is conserved during this process, and therefore no external heat exchange is required for envelope re-equilibration.

TFM24 instead showed that the mass-loss rate in core-powered mass loss is energy-limited by the radiation in the atmosphere, through either incident bolometric irradiation or the outgoing thermal radiation associated with interior cooling, rather than by the core energy absorbed in the envelope as argued by \citet{Ginzburg16}. This leads to a characteristic mass-loss timescale of a few Myr, substantially shorter than previous estimates. Consequently, the outflow traditionally referred to as core-powered mass loss is expected to operate primarily shortly after disk dispersal \citep{Ikoma12,owenwu16}, when the rapid decrease in external confining pressure allows the extended atmosphere to expand and develop a strong pressure gradient. This rapid atmospheric escape is commonly referred to as \emph{boil-off}.

From this perspective, boil-off and core-powered mass loss do not represent fundamentally distinct physical processes. Rather, the distinction concerns the assumed source of energy for envelope re-equilibration. For boil-off, previous studies have attributed this energy to gravitational energy released during envelope contraction \citep{owenwu16} or to the cooling of the planet \citep{Rogers24}, hereby assuming the wind as not being energy-limited. However, these interpretations do not explicitly identify and model the physical origin of these argued processes. In contrast, TFM24 argued that the envelope does not require additional contraction or cooling to supply this energy, but instead identified the envelope's thermal energy as the energy source. The apparent discrepancy arises from the interpretation of the gravitational energy term: the increase in gravitational potential energy associated with lifting gas is an energy demand of the outflow as suggested by TFM24, rather than an energy source as interpreted by \citet{owenwu16}. 

In this context, an important open question is what fraction of the thermal energy used to lift material within the envelope originates from the envelope's primordial thermal energy acquired during formation and what fraction is supplied by the core through its luminosity. This determines the extent to which core luminosity enhances atmospheric mass loss and, consequently, whether it is appropriate to characterize the resulting outflow as \emph{core-powered mass loss}.

To date, no previous study of core-powered mass loss has directly evaluated the impact of core luminosity on the planetary population, or even on individual planetary radius evolution for its energy analysis. TFM24 provided an indication that this effect may be small. During the boil-off phase, the mass-loss timescale is much shorter than the thermal contraction timescale of the planet. Although core luminosity can contribute substantially to the total cooling luminosity, the amount of energy supplied by the core over a single mass-loss timescale is small compared with the thermal energy already stored in the envelope. Consequently, core luminosity has little time to affect the planetary radius or mass-loss rate during the rapid boil-off phase. In other words, for core luminosity to substantially enhance mass loss, the planet must remain sufficiently extended for a sufficiently long time for the cumulative effect of core heating to become important.

TFM24 further showed that the role of core luminosity may have been overestimated in analytical models of core-powered mass loss \citep[][hereafter, the analytical models]{Ginzburg18,Gupta19,Misener21}. These models overestimated the mass-loss timescale, thereby predicting prolonged coupling between core cooling and atmospheric mass loss. In addition, the energy treatment adopted in the analytical models does not conserve the energy of the remaining planet when mass is lost. As a result, the associated delay in planetary contraction can be incorrectly attributed to core luminosity, further exaggerating its inferred role in driving atmospheric escape.

Motivated by these considerations, we build upon TFM24 and investigate the role of core luminosity at the population level. Before assessing whether the outflow traditionally referred to as core-powered mass loss is substantially affected by core luminosity, we use the term \emph{bolometric-driven escape} to encompass both core-powered mass loss and boil-off, without distinguishing between the two. 

In addition, we construct a population synthesis model with improved treatments of several key physical processes and evaluate their impacts on the resulting population. Previous population studies \citep{Jin14,Chen16,Owen17,Ginzburg18,Gupta19,Rogers20} have combined thermal evolution with either photoevaporation or core-powered mass loss, typically assuming pure H$_2$ or H/He envelopes. Here, we couple bolometric-driven escape and photoevaporation self-consistently within unified population framework, adopt initial envelope mass fractions informed by planet formation models, and account for envelope metallicity. To facilitate comparison with these previous models, we consider sub-Neptunes with H/He-dominated envelopes fully mixed with metals overlying rock/iron cores, without including distinct water layers \citep{Seager10,lopez12,Aguichine25}. We also examine the effect of the radiative atmosphere, which can contribute substantially to the observed radii of sub-Neptunes \citep{Tang25} but is neglected or treated with simplified prescriptions in the aforementioned population models.

The paper is organized as follows. In Section \ref{sec:methods}, we describe the evolution model and population synthesis framework. In Section \ref{sec:results}, we present the effects of core luminosity and other key physical choices on the resulting planetary population. In Section \ref{sec:disc}, we discuss the implications of our results.

\section{Model Description} \label{sec:methods}
\subsection{Structure and Evolution Modeling} \label{subsec:evo}
The radius evolution of sub-Neptunes is primarily governed by two key processes, thermal contraction and mass loss, which we jointly model using the sub-Neptune evolution framework SEAMIST \citep{Tang26}.

Thermal contraction results from the increasing density of the envelope as it cools. The envelope's net cooling is strongly influenced by heating from the underlying rock/iron core. The rate at which this heat is supplied depends sensitively on the evolution of the rock/iron core \citep{Vazan18a}, including both its solidification, which regulates the efficiency with which energy is transported out of the core \citep{Tang25}, and the release of silicate vapor, which may inhibit heat transport from the core by suppressing convection in the envelope \citep{Misener22,Tejada26}.   

On the other hand, the planetary radius decreases as mass loss reduces the size of the envelope. SEAMIST self-consistently models mass loss from both bolometric-driven winds and XUV-driven photoevaporation, including the transition between these two regimes. Bolometric-driven escape dominates the mass loss at young ages \citep{Tang24} when the planet is large in radius \citep{Owen24}. The outflow transitions to photoevaporation when EUV radiation begins to penetrate to the isothermal sonic point of the bolometric-driven outflow. This transition typically occurs after $\sim$1-3 Myr, depending on the degree to which boil-off is energy-limited. We model boil-off using the energy-limited formulation of TFM24, who showed that the escape can be limited either by stellar bolometric heating deposited in the radiative zone or by the intrinsic thermal energy released by the planet in the deep radiative zone, where bolometric heating becomes inefficient:
\begin{equation}
\dot{M} = \min(\dot{M}_{\rm parker},\frac{L_{\rm bol}}{G M_p/R_{\rm rcb}},\frac{L_{\rm int}}{GM_p/R_{\rm rcb} - GM_p/R_{\rm \tau_h}})
\end{equation}
where $\dot{M}_{\rm parker}$ is the non-energy-limited Parker wind escape rate. The second term, $L_{\rm bol}/(G M_p/R_{\rm rcb})$, is the bolometric-limited escape rate, where $L_{\rm bol}$ is the planet's incident bolometric luminosity, $R_{\rm rcb}$ is the radius of the radiative-convective boundary defining the surface of the envelope, and $M_p$ is the total planetary mass. The third term, $L_{\rm int}/(GM_p/R_{\rm rcb} - GM_p/R_{\rm \tau_h})$, is the intrinsic-luminosity-limited escape rate, where $L_{\rm int}$ is the intrinsic luminosity and $R_{\rm \tau_h}$ is the radius below which the atmosphere becomes optically thick to bolometric radiation.

Following \citet{Owen24,Tang24}, we model the transition from bolometric-driven escape to photoevaporation by tracking the EUV optical depth at the sonic radius $r_s$:
\begin{equation}
\label{tau}
\tau_{\rm s} = \sigma_{\rm \nu0} H n_0 \Big|_s \sim 1
\end{equation}
where $\sigma_{\nu0}$ is the EUV absorption cross section of hydrogen at the absorption location, $H$ is the local scale height, $n_0$ is the local neutral hydrogen number density, and subscript “s” denotes the sonic point. Alternatively, TFM24 proposed that bolometric-driven escape becomes inefficient when the bolometric-driven mass-loss timescale, $t_{\rm loss}=M_{\rm env}/\dot{M}$, exceeds the envelope cooling timescale, $t_{\rm cool}=E_{\rm tot}/(L_{\rm int}-L_{\rm core})$,
owing to the exponential decline of the mass-loss rate with decreasing planetary radius. Here, $E_{\rm tot}$ is the total envelope energy, including both gravitational and thermal energy, and $L_{\rm core}$ is the core luminosity. 

We have tested population calculations using both transition criteria, as well as a fixed transition time of 5 Myr, and find minimal differences with mean-square percentage variation of $\sim$1\% in the final radius distributions. This is because both photoevaporative and bolometric-driven mass-loss timescales are longer than $\sim$10 Myr at these young ages as the planetary radius decreases during boil-off, such that delaying the transition would produce only a negligible additional loss of envelope mass. 
Our population results are therefore insensitive to the simplified transition criterion adopted in Eq. \ref{tau} and to the hydrodynamic details of the transition between the two outflow regimes. 

After the transition, the outflow becomes XUV-driven. \citet{Tang25b} showed that, at young ages when the planet remains inflated following boil-off, the XUV-driven outflow is governed by a balance between XUV absorption and the conversion of absorbed energy into thermal and kinetic energy. This regime, termed thermal-energy-mediated photoevaporation (TEMP), subsequently transitions into the classical energy-limited regime \citep{Watson81,lopez12} when the gravitational energy dominates over the thermal energy. We model TEMP, energy-limited photoevaporation, and their transition following the analytical fit to the self-consistent hydrodynamic calculations of \citet{Tang25b}:
\begin{equation}
\label{mdot-scaling}
\dot{M} = 
\begin{dcases}
    \frac{\eta F_{\rm{EUV}}R_{\rm{base}}^3}{GM_p},     & \text{if } \lambda_s \geq 1\\
    \frac{\epsilon F_{\rm{EUV}} R_{\rm{base}}^2 \mu_s}{kT_s} ,     & \text{if } \lambda_s < 1\\
\end{dcases}
\end{equation}
The energy-limited efficiency $\eta$ has been shown to vary substantially among planets with different gravitational potentials \citep{Salz16} and atmospheric metallicities \citep{Owen18,Huang23,Broome25}. For pure H/He atmospheres, $\eta$ can reach values as high as $\sim$0.3-0.4 \citep{Tang25b} after the TEMP phase, with earlier work suggesting a range of 0.1-0.3 \citep{Lammer09}, whereas metal-line cooling may reduce $\eta$ by more than a factor of five at high metallicities \citep{Kubyshkina26}. However, the dependence of $\eta$ on metallicity remains poorly constrained. The assumed energy range of photoionizing photons can also affect $\eta$; the higher values reported by \citet{Tang25b} may in part result from their inclusion of X-rays. In addition, reported values of $\eta$ vary substantially in the literature because different studies adopt different definitions of the characteristic radius in Eq. \ref{mdot-scaling}. Given these uncertainties, we adopt a constant $\eta=0.1$ \citep{lopez12}. We derive $\epsilon$, the sonic-point temperature $T_s$, and the mean molecular weight $\mu_s$ following the prescription of \citet{Tang25b}. The transition between the two regimes is determined by the atmospheric escape parameter evaluated at the sonic point:
\begin{equation}
\lambda_s = (GM_p/R_s)/(kT_s/\mu_s)
\end{equation}

SEAMIST computes the 1D planetary structure at each evolutionary timestep. From the bottom up, the structure consists of an iron core, a silicate mantle, a convective envelope composed of primordial H/He, and a radiative atmosphere. SEAMIST self-consistently models the thermal evolution and energy transport of the rock/iron core, including its solidification and associated energy release. The equations of state (EOS) for the envelope are computed using an additive-volume law with a correction for the entropy of mixing. The baseline EOSs for H, He, and H$_2$O are sourced from \citet{Chabrier19,Mazevet19}. The radiative atmosphere is modeled as either hydrostatic or steady-state, depending on the evolutionary stage and altitude, i.e., on whether the atmosphere is undergoing an outflow. We do not consider volatile exchange at the magma--envelope interface in our models; its potential role is discussed in Section \ref{subsec:volatile}.

We consider atmospheric metallicities ranging from solar to 300$\times$ solar and assume that the envelope metallicity is equal to the atmospheric metallicity. Metallicity affects radius evolution through several mechanisms, including changes to the interior density, atmospheric scale height, bolometric-driven and photoevaporative mass-loss rates, envelope cooling, and the atmospheric temperature-pressure (T--P) profile. All of these effects except the latter have been discussed and implemented in \citet{Tang25b,Tang26}. Because the radiative atmosphere can contribute substantially to the observed planetary radius, as demonstrated by \citet{Tang25}, variations in the \emph{T--P} profile can significantly alter the atmospheric scale height and therefore the planetary radius. To account for this effect, we fit the \emph{T--P} profiles of the grid of 1D radiative-convective equilibrium atmospheres, computed over a large range of surface gravity and intrinsic fluxes, from \citet{Tang25} with the analytical gray-atmosphere model of \citet{Guillot10}. The \citet{Guillot10} model has two free parameters, the opacities to bolometric and thermal radiation. We find that metallicity primarily affects the T-P profiles through the thermal opacity, whereas the bolometric opacity is primarily determined by the irradiation temperature. We provide the best-fit parameters in Table \ref{tab:op}.

\begin{deluxetable}{lcccccccccccc} \tablecaption{Best-fit opacities \label{tab:op}} 
\tabletypesize{\scriptsize}
\startdata \tablehead{ \colhead{Metallicity (times solar)} & \colhead{1} & \colhead{3} & \colhead{10} & \colhead{30} & \colhead{100} & \colhead{300}} 
$\kappa_{\rm th}$ & 0.02 & 0.025 & 0.035 & 0.045 & 0.06 & 0.08\\ 
\hline
\hline
\colhead{Irradiation temperature (K)} &
\colhead{390} &
\colhead{631} &
\colhead{775} &
\colhead{919} &
\colhead{1069} &
\colhead{1222} &
\colhead{1379} &
\colhead{1540} &
\colhead{1706} &
\colhead{2049} &
\colhead{2227} &
\colhead{2410} \\
\hline
$\kappa_{\rm v}$ & 0.0001 & 0.00035 & 0.00068 & 0.0010 & 0.0014 & 0.0019 & 0.0026 & 0.0034 & 0.0039 & 0.0040 & 0.0043 & 0.0060 \\ 
\enddata 
\tablecomments{Thermal opacity $\kappa_{\rm th}$ and optical opacity $\kappa_{\rm v}$ adopted in the \citet{Guillot10} atmospheric models, calibrated to reproduce the numerical atmosphere models.}
\end{deluxetable}

\subsection{Population Modeling} \label{subsec:pop}
We construct synthetic population models using SEAMIST, each consisting of $N=1000$ model planets, with the planetary parameters randomly drawn from their underlying distributions that we formulate below. These parameters include the rock/iron core mass $M_{\rm ric}$, orbital period $P$, and stellar mass $M_s$. 

Previous sub-Neptune population studies have often relied on tabulated mass-radius relationships \citep{Chen16,He26} or simplified planetary evolution using analytical models \citep{Owen17,Ginzburg18,Rogers20}, to reduce the computation time. Here, we instead construct synthetic populations using fully self-consistent numerical evolution calculations starting from the epoch of disk dispersal. 
We evolve each planet to 10 Gyr  and track the resulting distributions of optical transit radius $R_p$ and envelope mass fraction $f_{\rm env}$.

To assess the potential influence of statistical fluctuations on our final distributions, we performed three validation population calculations in which the planetary parameters were independently redrawn from the same probability distributions adopted in our fiducial model (described in Section \ref{sec:results}). We find a mean-square percentage variation in the planetary occurrence rate of 4\%. This is substantially smaller than the observational uncertainties in \citet{Fulton18}, indicating that our predicted radius distributions are insensitive to the choice of $N$. Indeed, our choice of $N=1000$ is comparable to the sample size of 907 planets remaining after applying the selection filters in \citet{Fulton18}.

The underlying distributions of ${M_{\rm ric},P,M_s}$ play a critical role in shaping the final planetary distributions and are chosen as follows. First, we calibrate the $M_s$ distribution for FGK stars within the parameter ranges targeted by classic studies of small-planet occurrence rates \citep{Fulton18,Kunimoto20,Dattilo23}. We fit the stellar-mass probability distribution using two stellar catalogs: one containing all confirmed Kepler planets from the NASA Exoplanet Archive \citep{NasaExoplanetArchive} and the other from \citet{Dattilo23}. Both catalogs yield consistent fits. The resulting distribution follows a normal distribution:
\begin{equation}
p(M_s) \propto
\exp\left[-\frac{(M_s-\bar{\mu}_s)^2}{2\sigma_s^2}\right],
\end{equation}
where $\bar{\mu}s=0.93M\odot$ and $\sigma_s=0.18M\odot$ are the mean and standard deviation of the distribution, respectively. Later-type stars emit a larger fraction of their total energy in the XUV \citep{Jackson12}, leading to more efficient photoevaporative mass loss \citep{Owen17}. While we find this effect to be important, we identify another effect that often has a larger impact on atmospheric mass loss. The longer pre-main-sequence contraction phase of low-mass stars results in much higher bolometric luminosities during the first $\sim$100 Myr than their present-day luminosities. The enhanced bolometric luminosity also leads to higher XUV luminosity and drives more rapid envelope loss through both boil-off and photoevaporation in late-type stars compared to early-type stars at young ages. An exception is F-type hosts, for which significant stellar radius inflation at ages of a few Gyr can enhance the photoevaporative mass-loss rate relative to G-type stars.

Our $P$ distribution is calibrated using catalogs corrected for observational completeness, following the same procedure adopted by \citet{Fulton18}. Planets on wider orbits have a lower geometric transit probability and are therefore underrepresented in transit surveys, requiring a higher weight when calculating the intrinsic occurrence rate. The $P$ distributions reported in the literature \citep{Howard12,Petigura18,Kunimoto20} are approximately flat in logarithmic period space at $P>10$ days but decline toward shorter periods. We fit the $P$ distribution using the updated scaling of \citet{Dattilo23}, who find that the approximately flat portion has a slight positive slope. Because their original scaling is provided for specific planet-radius bins, we refit the scaling for planets with radii in the range 1-4$R_\oplus$. This gives
\begin{equation}
p(P) \propto P^{\beta-1}\{1-\exp{[-(P/P_0)^\gamma]}\}
\end{equation}
where we find $P_0=7$ days, $\beta=0.21$, and $\gamma=2.0$. To remain consistent with \citet{Fulton18}, we restrict the orbital period to $P<100$ days. For each individual planet in a population run, we use its orbital period and stellar mass to calculate the incident bolometric and XUV fluxes as a function of stellar age, following the stellar evolution model of \citet{Johnstone21}. These incident fluxes are crucial in determining the planet's radius evolution through their effects on the mass-loss rate, cooling rate, and atmospheric structure.

Yet, the most uncertain of these distributions is the core-mass distribution, $p(M_{\rm ric})$, for several reasons. First, the masses of relatively few sub-Neptunes have been measured. Moreover, these measurements are predominantly obtained through transit-timing variations (TTVs), which are biased toward more massive planets and depend on assumptions about the orbital configuration. Second, the rock/iron mass and the corresponding envelope mass, $M_{\rm env}=M_p-M_{\rm ric}$, are degenerate for a given observed planetary mass $M_p$, because planets with different combinations of ${M_{\rm env},M_{\rm ric}}$ can produce similar radii while having different interior entropies and atmospheric metallicities, neither of which is well constrained observationally. Consequently, we treat $p(M_{\rm ric})$ as a free model input parameter and explore a range of possible probability functions. The first one is the Rayleigh distribution, as adopted in \citet{Owen17,Ginzburg18,Wu19}:
\begin{equation}
\label{rayleigh}
    p(M_{\rm ric}) \propto
M_{\rm ric}
\exp\left(-\frac{M_{\rm ric}^2}{2\sigma_{\rm ric}^2}\right)
\end{equation}
In this work, we explore a range of values, $\sigma_{\rm ric}=[3,4,5]M_\oplus$. A simple fit to the observed total mass distribution yields the following functional form:
\begin{equation}
\label{best-pcore}
p(M_{\rm ric}) \propto
\begin{dcases}
    \rm constant ,     & \text{if } \sigma_{\rm ric} < 9 M_\oplus\\
    M_{\rm ric}\exp\left(-\frac{M_{\rm ric}^2}{2\sigma_{\rm ric}^2}\right),     & \text{if } \sigma_{\rm ric} \geq 9 M_\oplus\\
\end{dcases}
\end{equation}
where in the above equation we find $M_{\rm ric}=5_\oplus$ to best fit the observation.

Given the degeneracy in the envelope mass of sub-Neptunes, previous population studies have often adopted initial envelope mass fractions drawn from either a log-normal distribution \citep{Wu19,Rogers20} or a log-uniform distribution \citep{Owen17}. These log-normal distributions narrowly peak at 3-4\%, substantially lower than the envelope mass fractions predicted by gas-accretion models \citep[$>10\%$,][]{Lee15,Ginzburg16,Mordasini20}. Moreover, these empirical distributions do not account for the expected dependence of the initial envelope mass on core mass, equilibrium temperature, and disk lifetime, which can critically affect the planet-nebula thermal equilibrium and, consequently, the initial gas inventory.

Motivated by these considerations, we adopt the gas-accretion model of \citet{Ginzburg16} as our default prescription for the initial envelope mass. We assume a disk lifetime of 5 Myr, following the observational trend of \citet{Mamajek09}. We then introduce a multiplicative factor $\zeta$ to scale the initial envelope mass relative to the default value, considering $\zeta=[30\%,50\%,100\%]$ as a free parameter when exploring the best-fit models. This approach allows us to capture uncertainties in the gas-accretion model. In particular, the origin of envelope metals remains uncertain: they may be acquired through processes such as icy pebble accretion or produced through chemical reactions between the envelope and the planetary interior. We therefore expect the initial envelope mass to deviate from the value predicted under the assumption of pure H accretion. The scaling factor $\zeta$ also provides a simple way to explore scenarios in which the initial envelope mass is reduced by processes not included in the gas-accretion model, such as impact-driven atmospheric erosion \citep{Inamdar16,Kegerreis2020}, which may potentially contributes to the population distribution \citep{He26}, or gas-poor formation pathways \citep{Lee14}.

\section{Results} \label{sec:results}
We run a grid of population models by varying the initial envelope mass factor $\zeta$ (3 grid points), the core-mass distribution (4 grid points), and the envelope metallicity, with values of $1$, $30$, $50$, $100$, and $300\times$ solar (5 grid points). We additionally toggle core luminosity and boil-off on and off to assess their importance for the resulting radius distribution. This amounts to a total of 180 population runs, comprising more than 100,000 model planets. We compare these models with the observations of \citet{Fulton18} and designate the best-fitting model as our fiducial model. The fiducial model adopts a Rayleigh core-mass distribution with $\sigma_{\rm ric}=3M_\oplus$, an envelope metallicity of $100\times$ solar, and $\zeta=50\%$, corresponding to a typical initial envelope mass fraction of $\sim10\%$.

\begin{figure}
\centering
\includegraphics[width=0.5\textwidth]{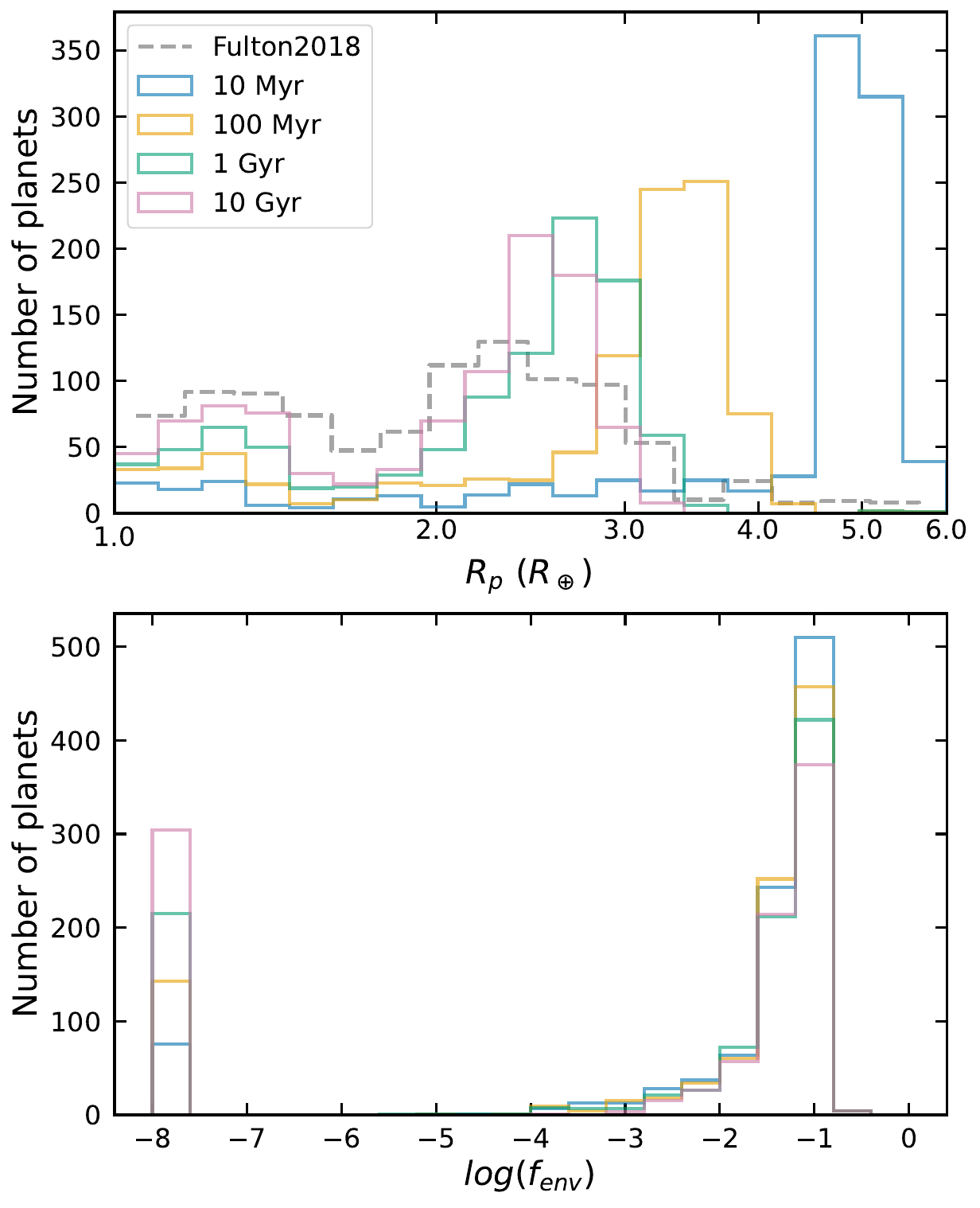} 
\caption{ Comparison of the distributions at different ages. The top and bottom panels show the optical transit radius and envelope mass fraction distributions, respectively. For comparison, the gray dashed line shows the observed radius distribution from \citet{Fulton18}.
}
\label{fig:age}
\end{figure}

In Figure \ref{fig:age}, we show the evolution of the radius and envelope mass fraction distributions for our fiducial model. The peak radius of the sub-Neptune population decreases over time as an increasing fraction of planets evolve into super-Earths (bottom panel). At 10 Myr, shortly after the boil-off phase, the sub-Neptune population has a typical radius of $\sim5R_\oplus$. On Gyr timescales, the variation in the radius distribution becomes minimal. In later sections, we adopt a default planetary age of 5 Gyr when not otherwise specified, consistent with the median age of Kepler objects.

Notably, we find a wide range of mass-loss timescales among sub-Neptunes, as reflected in the continuous growth of the super-Earth population over time. Previous observations of sub-Neptune outflows \citep{Berger20,david21,Christiansen23} have commonly adopted photoevaporation timescales of $\sim100$ Myr and have often used this timescale to distinguish photoevaporation from core-powered mass loss, which is typically assumed to operate over timescales of several Gyr following \citet{Ginzburg18,Gupta19}. Our results suggest that such a criterion can lead to a misleading interpretation of the physical origin of observed outflows.

In the following subsections, we detail the effects of each input parameter that we find important, examine the roles of core luminosity and atmospheric escape, discuss the contribution of the radiative atmosphere and boil-off, and compare the relative importance of boil-off and photoevaporation.

\subsection{Role of Core-Mass Distribution} \label{subsec:pcore}
\begin{figure}
\centering
\includegraphics[width=0.5\textwidth]{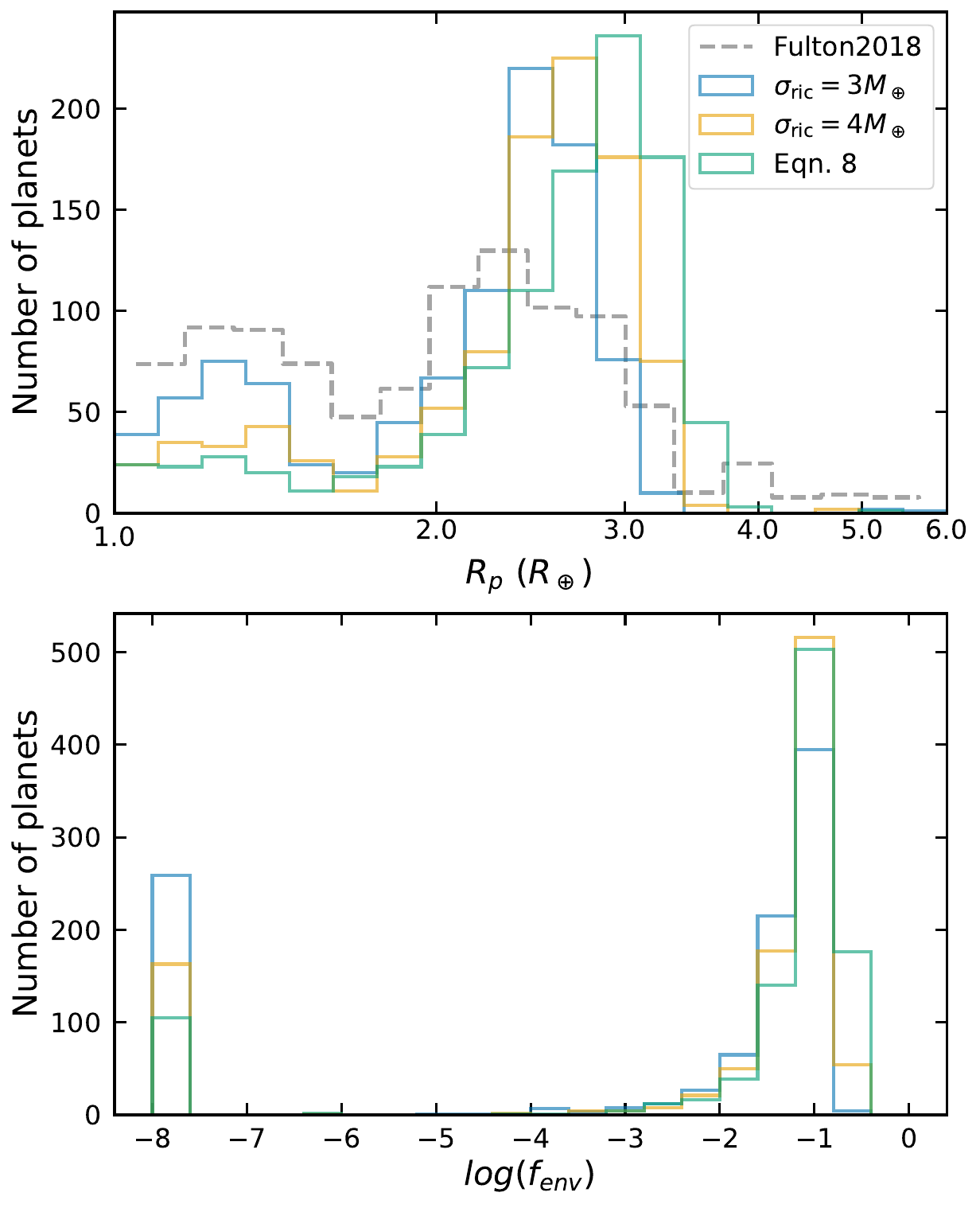} 
\caption{ Comparison of population models with different core-mass probability density functions. We adopt a planetary age of 5 Gyr. The blue and orange models adopt the Rayleigh distribution given by Eqn. \ref{rayleigh}, while the green model adopts the distribution given by Eqn. \ref{best-pcore}. The green model provides the best fit to the observed planetary mass distribution. The observed radius distribution from \citet{Fulton18} is shown by the gray dashed line.
}
\label{fig:pcore}
\end{figure}

In Figure \ref{fig:pcore}, we present simulated planetary occurrence distributions for populations with different core-mass probability functions. We include the observations of \citet{Fulton18} for comparison (gray dashed). All other input parameters remain the same as our fiducial model. In general, we find that the Rayleigh distribution with $\sigma_{\rm ric}=3M_\oplus$ (blue) provides the best match to the observed bimodal distribution, consistent with the findings of \citet{Owen17}, although our physical assumptions differ substantially (e.g. metallicity). Increasing $\sigma_{\rm ric}$ shifts both the sub-Neptune peak and the radius cutoff toward larger radii, thereby broadening the radius gap, while also significantly reducing the super-Earth occurrence rate, as illustrated by the $\sigma_{\rm ric}=4M_\oplus$ model (orange). We find that the core-mass distribution given by Eqn.\ref{best-pcore} generally reproduces the observed planetary mass distribution, but produces a radius distribution that differs significantly from the observations. This suggests that the underlying  planetary mass distribution may be more concentrated around lower masses $\sim3M_\oplus$ than implied by the observed planetary mass distribution, whose mean is $\sim5M_\oplus$.

\subsection{Role of Envelope Metallicity} \label{subsec:met}
\begin{figure}
\centering
\includegraphics[width=0.5\textwidth]{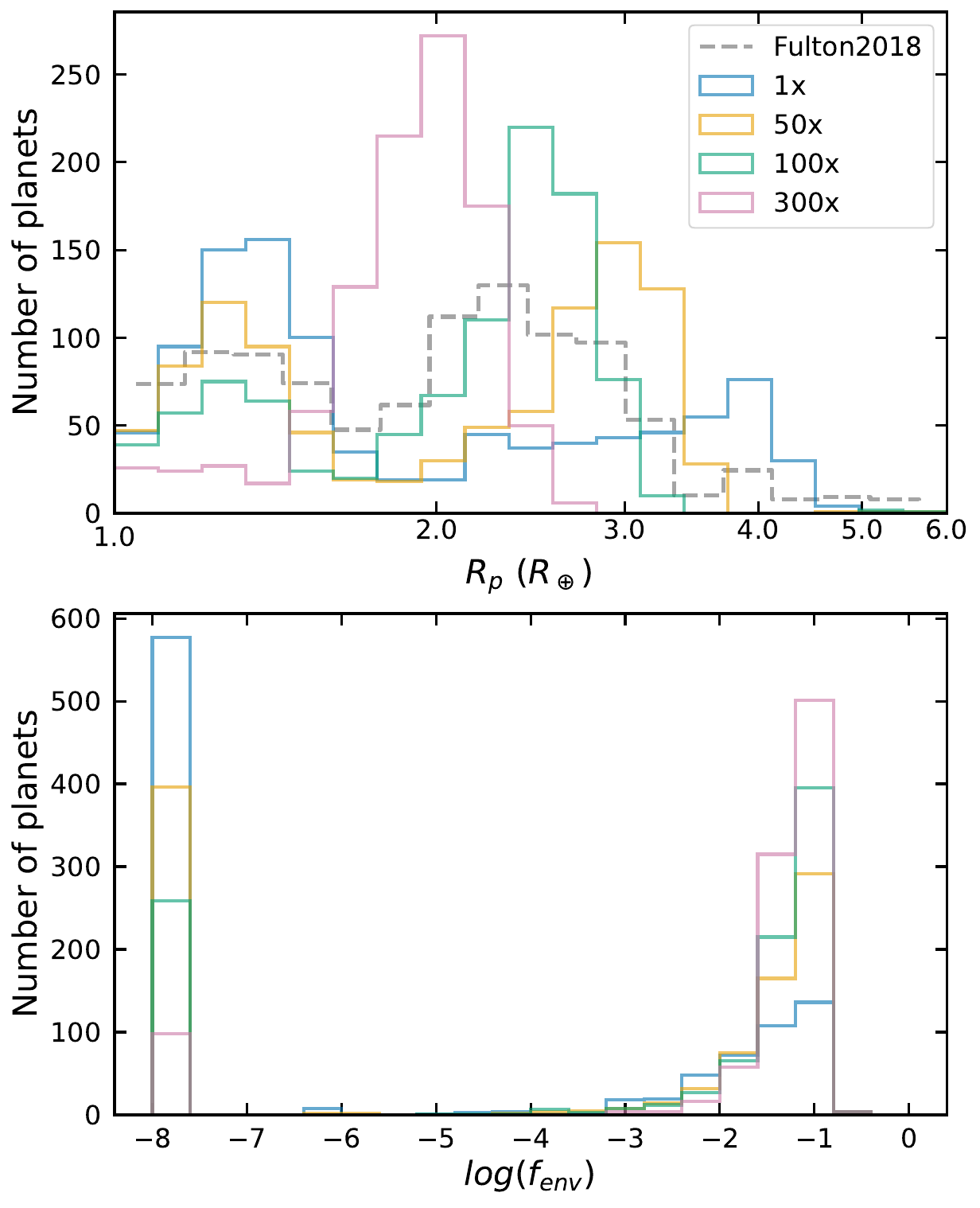} 
\caption{ Comparison of population models with different envelope metallicities. We adopt a core-mass distribution with $\sigma_{\rm ric}=3,M_\oplus$, and the initial envelope mass fraction is scaled to $\zeta=0.5$, corresponding to a $50\%$ reduction from the default value.
}
\label{fig:met}
\end{figure}

In Figure \ref{fig:met}, we present the radius distributions (top) and corresponding envelope mass fraction distributions (bottom) for populations with different envelope metallicities. All other input parameters are fixed to those of our fiducial model. In general, we find a larger fraction of super-Earths at lower metallicities, as lower-metallicity envelopes are more susceptible to both boil-off and photoevaporative mass loss, as discussed in \citet{Tang25b}. Increasing the envelope metallicity also reduces the planetary radius by increasing the interior density and decreasing the atmospheric scale height. This shifts both the sub-Neptune population peak and the radius cliff---the boundary characterized by a sharp decline in the occurrence rate---to smaller radii. These shifts narrow the radius gap, while having minimal impact on its depth.

Interestingly, no distinct radius gap is present in the $300\times$ solar model (magenta), which has a metal mass fraction of $Z\sim0.8$, corresponding to a water-dominated envelope. In the low-metallicity models ($\lesssim30\times$ solar), a significant portion of planets completely lose their envelope by boil-off, effectively placing the transition from boil-off to photoevaporation within the super-Earth regime.

Previous sub-Neptune population studies have generally assumed solar metallicity or pure H/He envelopes \citep{Chen16,Owen17,Jin18,Rogers20}. However, across the population grid explored here, the solar-metallicity models cannot reproduce a bimodal distribution consistent with the observations as they did. Instead, the solar-metallicity model (blue) in Figure \ref{fig:met} produces a broad and diffuse sub-Neptune population rather than a distinct bimodal distribution. This behavior is consistent across all of the $1\times$ solar models we explore. In addition, the radius gap is substantially wider than observed: the narrowest gap among our solar-metallicity models spans $1.7-2.5R_\oplus$, occurring in the models with the smallest initial envelope masses. The discrepancy between our results and previous studies likely arises from their simplified treatment or omission of the radiative atmosphere and their neglect of boil-off, as we demonstrate in later sections.

Overall, we find that the $100\times$ solar models (green) best reproduce the observed width of the radius gap, whereas the $50\times$ solar models (orange) provide the best match to the relative occurrence of the sub-Neptune and super-Earth populations. Our models predict a preferred envelope metallicity range of $50-100\times$ solar, which provides a testable prediction for future observations with JWST transmission spectroscopy.

\subsection{Role of Initial Envelope Mass} \label{subsec:env}
\begin{figure}
\centering
\includegraphics[width=0.5\textwidth]{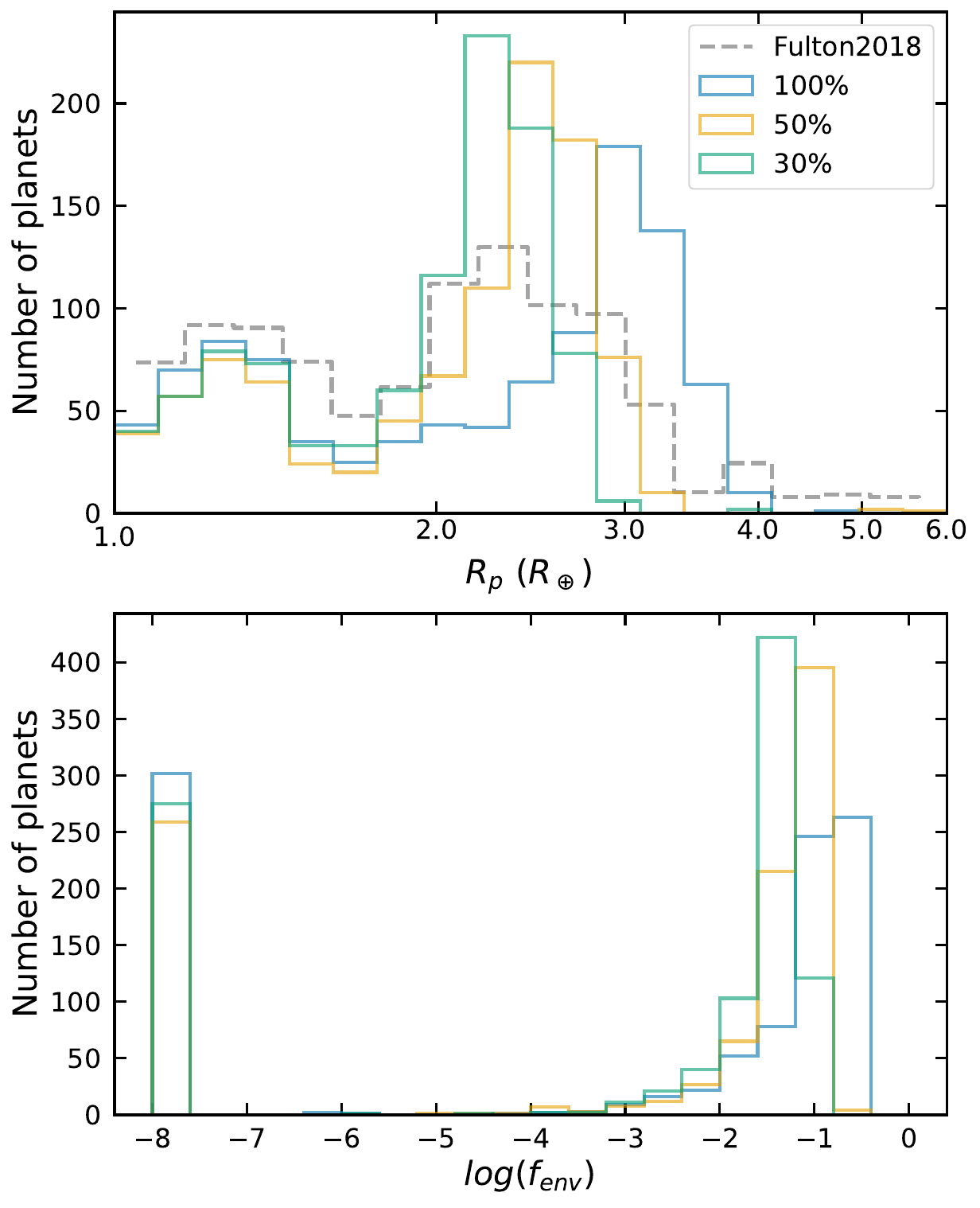} 
\caption{ Comparison of population models with different envelope metallicities. The envelope metallicity is consistent with our fiducial model (100x solar).
}
\label{fig:frac}
\end{figure}

In Figure \ref{fig:frac}, we present models with different initial envelope mass fractions, scaled by the multiplicative factor $\zeta$. We find that reducing the initial envelope mass from the default value to $30\%$ of the default value has little impact on the relative occurrence of sub-Neptunes and super-Earths, as seen in the height of the super-Earth peak on the left. This is expected for two reasons. First, planets that are transformed into super-Earths have all undergone a boil-off phase. Second, \citet{Tang24} demonstrated that the envelope mass fraction remaining after boil-off is largely insensitive to the initial envelope mass, owing to the decoupling between thermal contraction and mass loss (which is further discussed in Section \ref{subsec:core}). This insensitivity is also evident in the population of thin-envelope planets, with $\log(f_{\rm env})$ between $-3$ and $-2$, shown in the bottom panel.

In contrast, decreasing the initial envelope mass primarily shifts the radius cutoff and the location of the sub-Neptune peak to smaller radii, narrowing the gap. This is because the radius cutoff is determined primarily by massive cores retaining sufficiently massive envelopes to remain relatively insensitive to atmospheric mass loss. Reducing the initial envelope mass therefore decreases the final radii and envelope mass of these planets, shifting the radius cutoff toward smaller radii. In addition, decreasing the initial envelope mass slightly reduces the depth of the radius gap by producing more planets with thin envelopes that populate the gap..

\subsection{Effect of Core Luminosity} 
\label{subsec:core} 
\subsubsection{Outflow Energetics} \label{subsubsec:energetics}
In this section, we examine the energy balance of the planet and assess how core luminosity affects the outflow by heating the envelope, thereby setting the lower boundary for both photoevaporative and bolometrically driven escape. Previous studies of sub-Neptune evolution have examined the role of core luminosity in atmospheric escape (often referred to as core-powered mass loss) using several different approaches, including comparisons of the total energy \citep{Ginzburg16,Gupta19,Misener21}, the cooling luminosity \citep{Rogers24}, and the relevant timescales \citep{Tang24}.

In the classical energy approach \citep{Ginzburg16,Gupta19,Misener21}, the total energy available from the core is compared with the gravitational binding energy of the envelope to estimate whether the core can supply sufficient energy to remove the entire envelope. However, the cooling energy of the core is not necessarily converted into useful work to unbind atmospheric gas. In a core-energy-dominated planet, the intrinsic luminosity of the envelope, $L_{\rm int}$, can be comparable to $L_{\rm core}$, indicating that a substantial fraction of the energy supplied by the core is transported through the envelope and ultimately radiated away at its surface, rather than being used directly to drive mass loss. Moreover, these calculations generally neglect the energy carried away by the escaping gas through advection. As demonstrated by TFM24, this advective energy loss means that the energy budget of the remaining planet is not conserved in the manner assumed by the classical energy argument, potentially leading to an overestimate of the role of core luminosity in atmospheric escape.

The luminosity-based approach of \citet{Rogers24} instead compares the core luminosity with the time derivative of the envelope's total energy, $\dot{E}_{\rm env,tot}$. While this provides a more direct comparison of the relevant energy fluxes, it does not fully capture the cumulative nature of the effect of core luminosity on mass loss. Core luminosity can enhance atmospheric escape by continuously supplying heat to the envelope, thereby delaying its contraction \citep[or causing thermal inflation as per][]{Tang25} and maintaining a larger planetary radius. The resulting enhancement of mass loss therefore depends on how long the core luminosity acts on the envelope. For example, even if $L_{\rm core}$ is orders of magnitude larger than $\dot{E}_{\rm env, tot}$, the planetary radius may change very little over a sufficiently short time interval, resulting in little immediate enhancement of the mass-loss rate. Furthermore, their energy budget includes both the change in gravitational energy from the deep interior to infinity and the thermal energy of the escaping material. In contrast, $L_{\rm core}$ only heats in the envelope, which sets the lower boundary of the outflow at the RCB, while the radiative zone is primarily heated by stellar irradiation. Consequently, the change in the total binding energy from the deep interior to infinity is not directly relevant to the energy supplied by the core. Moreover, the energy demand, $\dot{E}_{\rm env, tot}$, should account only for the change in gravitational energy, rather than the total energy, because the thermal energy of the envelope serves as an energy supply that contributes to powering the outflow rather than constituting energy demand \citep{Tang24}.

In \citet{Tang24}, the role of core luminosity is instead characterized through a heating timescale,
\begin{equation}
t_{\rm heat} = \frac{U}{L_{\rm core}},
\end{equation}
which measures the timescale over which core luminosity can significantly delay the contraction of the planet, and is compared with the mass-loss timescale, $t_{\rm loss}$. This approach accounts for the cumulative nature of core heating while retaining the physical interpretation of the luminosity. These zeroth-order estimates show that during the boil-off phase, the mass-loss timescale is much shorter than the thermal evolution timescale of the core and envelope. Consequently, over the short timescale of mass loss, core luminosity produces only a negligible delay in planetary contraction, suggesting that it has a minor effect on the mass-loss rate during boil-off.

Motivated by the limitations of these previous approaches, we develop a refined method to quantify the role of core luminosity in atmospheric escape. We define an energy enhancement factor, $\phi$, as the ratio between the thermal energy supplied by the core and the gravitational energy required to lift material within the envelope over a time interval $\Delta t$:
\begin{equation}
\phi =
\frac{\displaystyle\int_{\Delta t}
L_{\rm core}
\frac{E_{\rm env,th}}{|E_{\rm env,tot}|}dt}
{\displaystyle\int_{\Delta t}
\dot{U}_{\rm loss}dt}.
\end{equation}
Here, $\Delta t$ encompasses either the boil-off and photoevaporation phases, which we treat as continuous mass-loss processes because, as demonstrated above, each phase should be considered as a whole when evaluating the cumulative effect of core luminosity. We define the total energy of the envelope as
\begin{equation}
\label{phi}
E_{\rm env,tot}=U_{\rm env}+E_{\rm env,th},
\end{equation}
where $U_{\rm env}<0$ is the gravitational binding energy and $E_{\rm env,th}$ is the thermal energy. The fraction of the core luminosity that is deposited as thermal energy available to the outflow (i.e. energy supply) is $E_{\rm env,th}/|E_{\rm env,tot}|$. $\dot{U}_{\rm loss}$ is a positive gain that reflects the energy demand in the envelope for material to be supplied at the RCB to the outflow during envelope re-equilibration (see TFM24). $\dot{U}_{\rm loss}$ is given by:
\begin{equation}
\dot{U}_{\rm loss} =
\int_{M_c}^{M_c+M_{\rm env}}
\left[
\frac{Gm(r)}{r}
-
\frac{Gm(r)}{R_{\rm rcb}}
\right]
\frac{\dot{M}}{M_{\rm env}}dm
\end{equation}
where we assume that the mass-loss rate of each envelope shell is proportional to its mass, such that a shell of mass $dm$ contributes a mass-loss rate of $\dot{M}dm/M_{\rm env}$. 

Physically, $\phi$ quantifies the additional thermal energy supplied by $L_{\rm core}$ relative to the gravitational energy required to re-equilibrate the envelope. As suggested by TFM24, in the absence of $L_{\rm core}$, this energy demand would instead be supplied by the planet's primordial thermal energy, and therefore reflects the amount of primordial thermal energy available to drive envelope expansion and mass loss. Consequently, $\phi$ measures the relative contribution of core-supplied thermal energy to the envelope's combined thermal-energy reservoir available for expansion.

Practically, we evaluate $\phi$ along an evolution track post facto. Thus, the effects of envelope cooling through $L_{\rm int}$ on the envelope's thermal structure and binding energy are already accounted for in the quantities entering the above equations. We note, however, that the additional heat supplied by $L_{\rm core}$ also enhances the cooling luminosity $L_{\rm int}$ relative to a model without core luminosity, thereby reducing the net thermal energy retained by the envelope. Because this effect cannot be directly quantified in a form that can be incorporated into the integral, we neglect it and therefore regard our estimated $\phi$ values as conservative upper limits. We have performed test calculations at several snapshots and find that this effect typically reduces the effective thermal-energy gain by $10-20\%$ for planets experiencing strong boil-off, but can be significant for planets not susceptible to mass loss.

\begin{figure}
\centering
\includegraphics[width=0.5\textwidth]{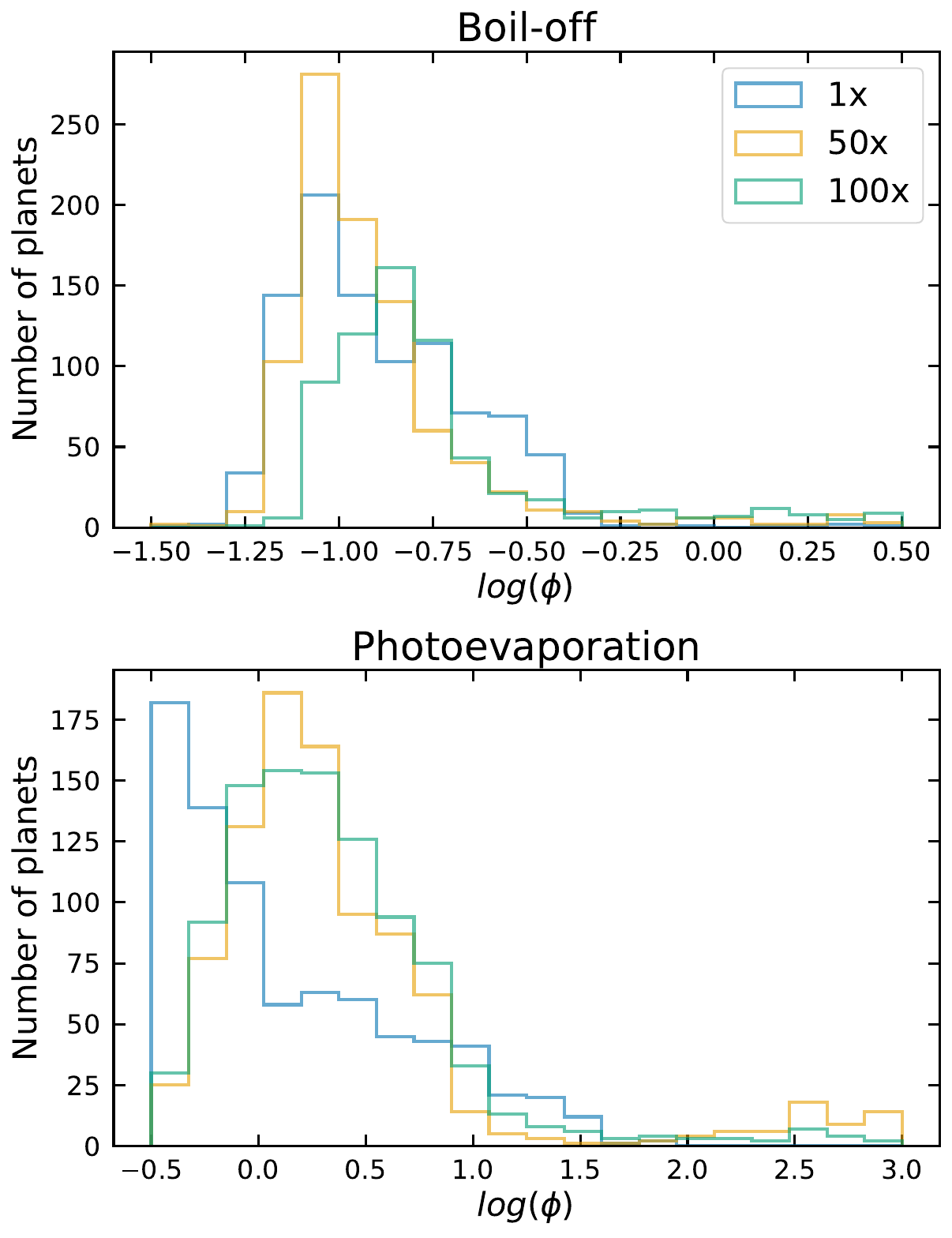} 
\caption{ Enhancement factor $\phi$ distributions for boil-off and photoevaporation. We show results for a range of envelope metallicities and adopt $\zeta=1$. We find that photoevaporation is significantly more dependent on core luminosity, whereas boil-off draws substantially less thermal energy from the core luminosity.
}
\label{fig:enhancement}
\end{figure}

In Figure \ref{fig:enhancement}, we show the enhancement factor during the boil-off (top) and photoevaporation (bottom) phases for a range of envelope metallicities and $\zeta=1$. We find that $\phi$ during boil-off typically peaks at $\sim10\%$, with the peak value increasing as the initial envelope mass is reduced because the slower mass-loss rate induces stronger coupling between mass loss and cooling. For our fiducial model, $\phi$ reaches a peak of $\approx20\%$. We find that $\phi$ increases with metallicity but never exceeds 25\% in all models. In contrast, $\phi$ during photoevaporation is typically one order of magnitude larger, reaching values of order unity. This indicates that photoevaporation is substantially more sensitive to core luminosity than boil-off, as the longer mass-loss timescale allows $L_{\rm core}$ to continuously supply energy to the envelope, leading to a stronger cumulative enhancement of mass loss over timescales ranging from $\sim100$ Myr to a few Gyr. This result is consistent with the predictions of \citet{Tang24}.

We emphasize that $\phi$ characterizes the fractional enhancement (i.e. how much additional fraction of mass loss may be driven by $L_{\rm core}$) of mass loss rather than its absolute magnitude, assuming the mass loss enhancement scales with its energy enhancement $\phi$. Despite the larger $\phi$ during photoevaporation, the absolute enhancement in mass loss is smaller, of 1wt\%, because the total mass lost through photoevaporation is much smaller than that lost during boil-off (see Section \ref{subsec:massloss}). 

The long tails toward high $\phi$ in both panels should be interpreted with caution, as Eqn.\ref{phi} is not applicable to these planets. These planets are not susceptible to mass loss thus holding thick envelopes and are instead dominated by intrinsic cooling, with $L_{\rm core}$ providing only a minor contribution to their thermal evolution. 

In general, we find that core luminosity constitutes a small fraction ($\sim10-20\%$) of the energy-budget in bolometric-driven escape (boil-off) but is significant in photoevaporation. 

\subsubsection{Radius Evolution} \label{subsubsec:radius}
\begin{figure*}
\centering
\includegraphics[width=0.9\textwidth]{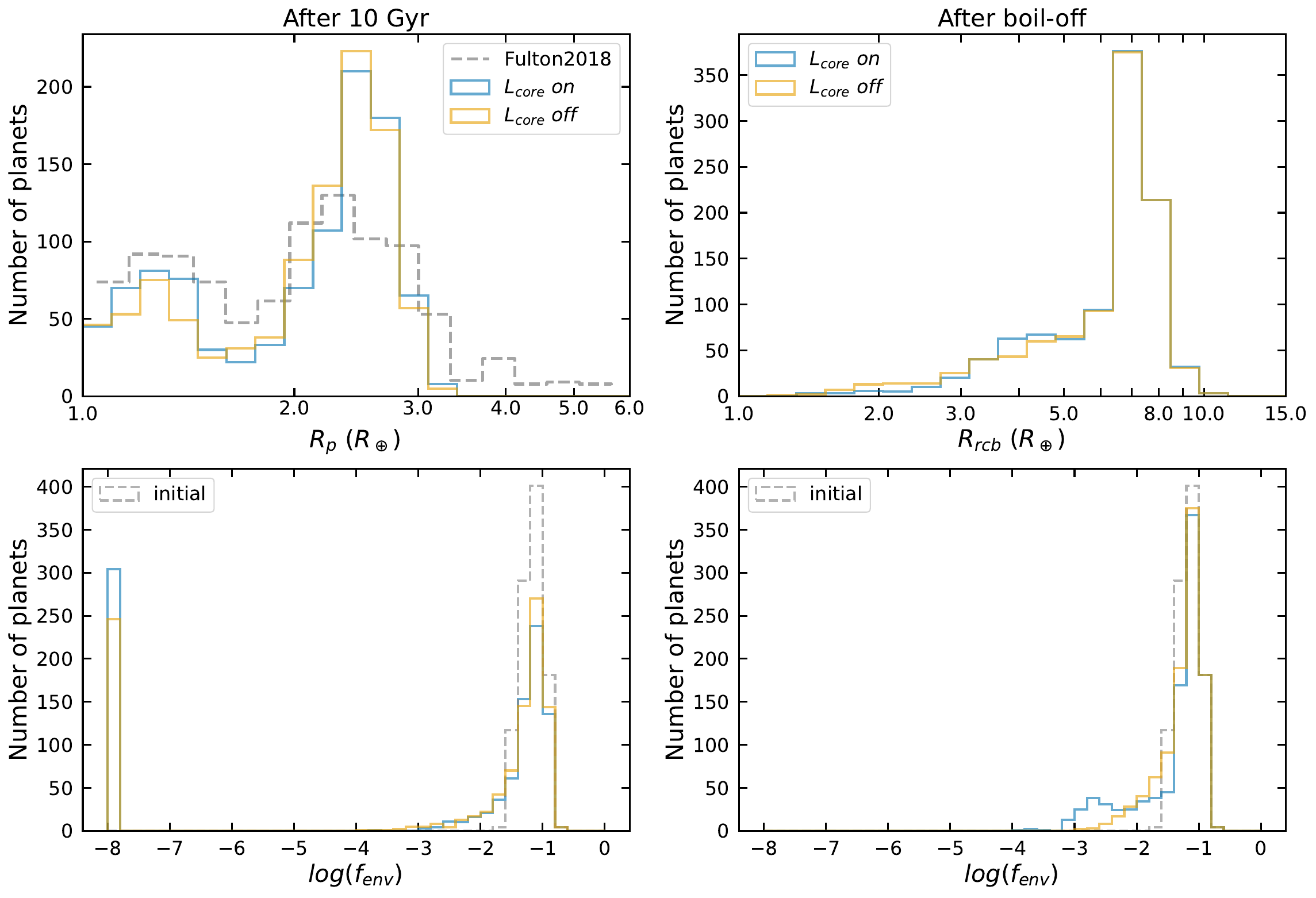} 
\caption{  Comparison of population models with core luminosity turned on (blue) and off (orange). The left and right panels show the distributions at 10 Gyr and at the boil-off-photoevaporation transition epoch, respectively. The initial envelope mass fraction distribution is shown by the gray dashed for comparison.
}
\label{fig:core}
\end{figure*}

We evaluate the role of core luminosity in planetary radius evolution and mass-loss enhancement by comparing models with $L_{\rm core}$ turned on and off. Figure \ref{fig:core} compares the radius (top) and envelope mass fraction (bottom) distributions for our fiducial model. The left panels show the distributions at 10 Gyr, while the right panels show the distributions at the epoch of the boil-off-photoevaporation transition. Note that in the right panels, we compare the envelope radius $R_{\rm rcb}$ rather than the transit radius $R_p$, as the radiative zone above the RCB has not yet reached its mature hydrostatic structure. In addition, $R_{\rm rcb}$ more directly reflects the impact of core luminosity on the envelope itself, and sets the lower boundary condition of the outflow.

We find that core luminosity has little impact on the envelope radius at the boil-off--photoevaporation transition (top right). The transition occurs at $R_{\rm rcb}/R_{\rm s}=0.05$-$0.1$, with an average value of $0.073$. Surprisingly, the differences in the transit-radius distributions also remain small even at 10 Gyr (top left). Although core luminosity enhances mass loss, the additional heating supplied to the envelope also maintains a larger radius, partially offsetting the radius reduction caused by the enhanced mass loss. Consequently, the net effect of core luminosity on long-term radius evolution is small.

In our fiducial model, we find an average absolute mass-loss enhancement of $0.6\mathrm{wt}\%$ by the end of boil-off, corresponding to a $24\%$ relative increase in the total mass lost, for planets that undergo strong boil-off accompanied by substantial thermal inflation. These planets typically lose $\gtrsim90\%$ of their initial envelope mass, as per \citet{Tang25}. When considering all planets that undergo a boil-off phase, the corresponding values decrease slightly to $0.4\mathrm{wt}\%$ and $23\%$, respectively. These results are consistent with the estimates presented in Section \ref{subsubsec:energetics}. 

We find that planets undergoing modest boil-off, with $\lesssim1\mathrm{wt}\%$ of their envelope mass lost (which constitute 25\% of all planets undergoing boil-off), can exhibit relative mass-loss enhancements exceeding $50\%$. This large relative enhancement arises from the stronger coupling between mass loss and thermal evolution for these planets, which places them close to the boil-off-photoevaporation transition boundary in radius space. Nevertheless, because the overall escape rate is low for these planets, the absolute enhancement in mass loss remains small.

After 10 Gyr, we find a $20\%$ higher occurrence of super-Earths in the population with $L_{\rm core}$ (bottom left), while the combined mass-loss enhancement amounts to an absolute increase of $1.1\mathrm{wt}\%$, corresponding to a relative increase of $38\%$. At 10 Gyr, photoevaporation primarily affects planets in the higher envelope-mass bins, as shown in the bottom left panel, whereas the bottom-right panel suggests that boil-off predominantly affects planets with lower envelope masses.

In the solar-metallicity models, the absolute enhancement increases while the relative enhancement decreases due to the faster mass-loss rate. The enhancement is $1\mathrm{wt}\%$ ($19\%$) for strong boil-off, $0.7\mathrm{wt}\%$ ($17\%$) for all planets undergoing boil-off, and $1.1\mathrm{wt}\%$ ($20\%$) when both boil-off and photoevaporation are included, where values in parentheses denote relative enhancements. The modest additional enhancement from photoevaporation reflects the smaller total mass loss from photoevaporation in solar-metallicity planets (see Section \ref{subsec:massloss}).

\subsection{Role of Radiative Atmosphere} \label{subsec:atmo}
Previous sub-Neptune population synthesis studies have either neglected the radiative atmosphere altogether \citep{Ginzburg18,Gupta19} or simplified it as a constant-scale-height layer by assuming an isothermal temperature profile and constant gravity \citep{Lopez13,Owen13,Chen16,Owen17,Rogers20}. However, given the low surface gravity and high irradiation of sub-Neptunes, their atmospheric scale heights can constitute a substantial fraction of the planetary radius. As demonstrated by \citet{Tang25}, the radiative atmosphere can contribute more than $30\%$ of the planetary radius. They further show that assuming constant gravity can substantially underestimate the radius because the scale height increases with altitude. This extended atmosphere is also sensitive to the temperature profile, with isothermal models underestimating temperatures below the transit altitude and thus the planetary radius. These effects become more pronounced at lower pressure levels, including the photoionization base $R_{\rm euv}$ \citep{RMC09}. Previous population studies often assume $R_{\rm euv}=R_p$ \citep{Owen13,Owen17,Jin18,Rogers20}. In contrast, recent models suggest that $R_{\rm euv}$ can exceed $2R_p$ \citep{Tang25b,Broome25} for low-gravity planets, potentially leading to substantial underestimates of photoevaporative mass-loss rates in previous models. 

\begin{figure}
\centering
\includegraphics[width=0.48\textwidth]{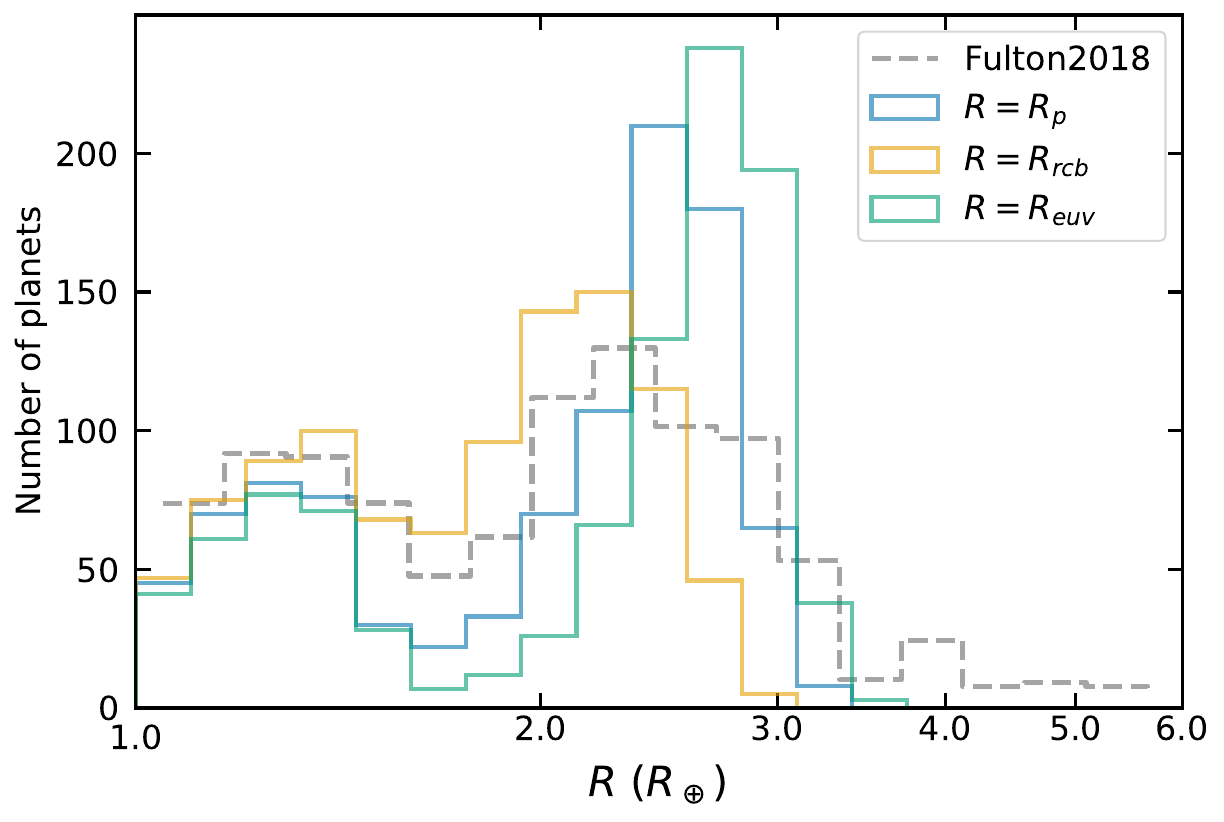} 
\caption{ Comparison of radius distributions evaluated at different atmospheric pressure levels for our fiducial model at 10 Gyr: the optical transit radius (blue), the envelope surface (orange), and the photoionization base (green).
}
\label{fig:atmo}
\end{figure}

Here, we examine the importance of the radiative atmosphere at the population level. In Figure \ref{fig:atmo}, we compare the radius distributions evaluated at three pressure levels: the RCB (envelope surface), the optical transit radius, and the photoionization base. For our fiducial model, the radius increases by an average of 15-20\% from $R_{\rm rcb}$ to $R_p$, which is significant given the relatively small observational uncertainties in planetary radius. This difference is smaller than that reported by \citet{Tang25}, owing to the higher mean molecular weight ($4.9$ amu) and older planetary age ($10$ Gyr) of our presented population. Interestingly, using the envelope radius provides a better match to the observations, which may partially explain the success of \citet{Ginzburg18} in reproducing the observed radius distribution.

The radius increases by an additional $\sim15\%$ from $R_p$ to $R_{\rm euv}$. This corresponds to an approximately $50\%$ increase in the energy-limited mass-loss rate, which is the dominant photoevaporation regime at 10 Gyr according to Eq. \ref{mdot-scaling}. We find these radius differences become substantially larger for low-metallicity planets, planets near the radius gap, and young planets.

\subsection{Boil-Off vs. Photoevaporation} \label{subsec:massloss}
Previous population studies have either neglected the boil-off phase while focusing on photoevaporation \citep{Lopez13,Owen13,Jin14,Chen16,Jin18,Rogers20} or adopted the classical analytical framework for core-powered mass loss \citep{Ginzburg18,Gupta19}. \citet{Tang24} later showed that this framework overestimates both the mass-loss timescale and the role of core luminosity, and that bolometrical-driven winds instead occur during a short-lived boil-off phase. In this section, we evaluate the role of boil-off relative to photoevaporation and assess how neglecting boil-off affects the resulting population.

\begin{figure}
\centering
\includegraphics[width=0.5\textwidth]{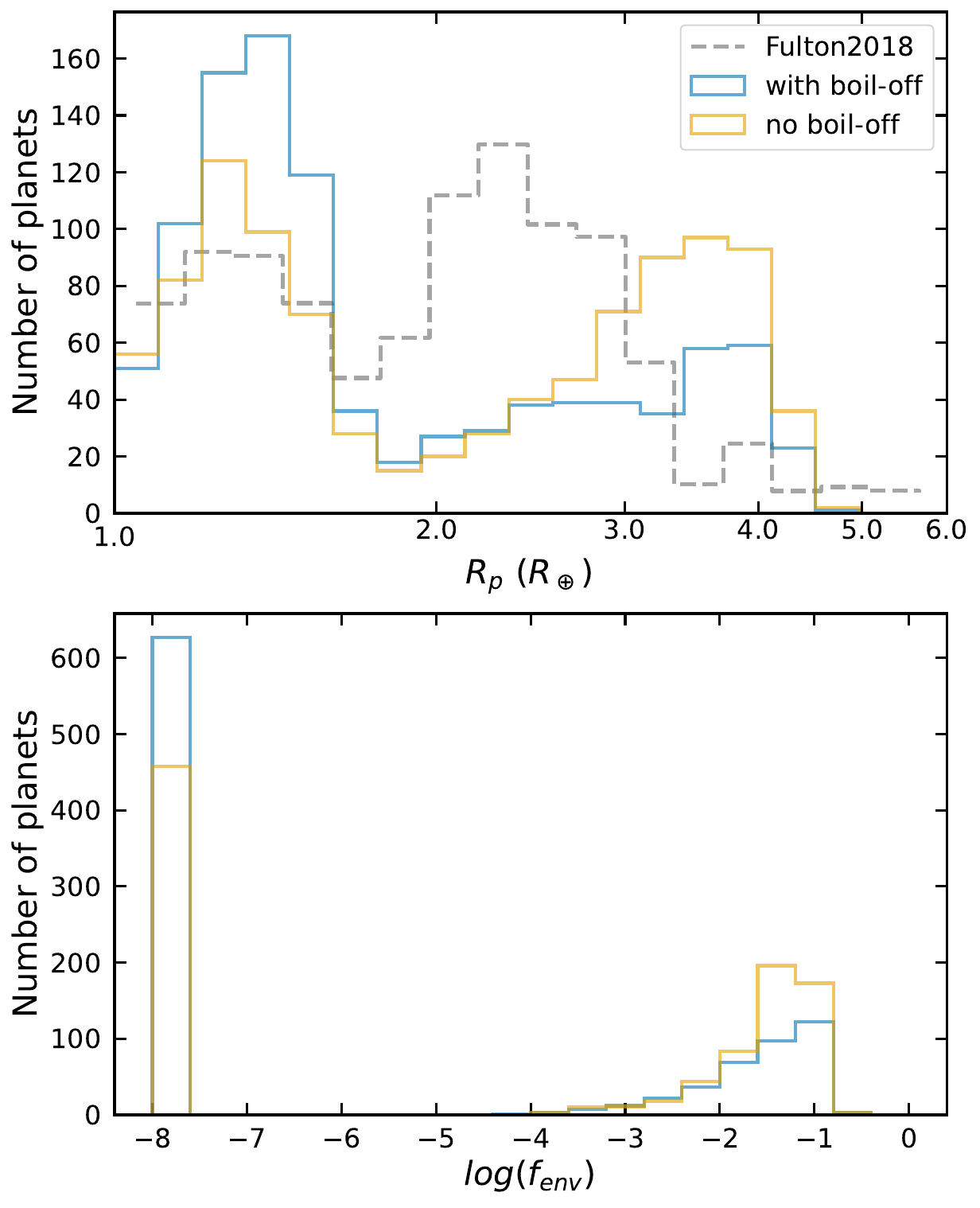} 
\caption{ Comparison of population models with (blue) and without (orange) boil-off. We adopt solar metallicity for consistency with previous studies.
}
\label{fig:boiloff}
\end{figure}

In Figure \ref{fig:boiloff}, we compare the radius distributions with (blue) and without (orange) a boil-off phase. To facilitate comparison with previous studies, we adopt solar metallicity consistent with their assumption. We find that neglecting boil-off can underestimate the super-Earth occurrence rate by up to $\sim50\%$. In contrast, it has little impact on the depth and location of the radius gap or the location of the radius cutoff, which are primarily determined by the initial envelope mass fraction and metallicity. Interestingly, neglecting boil-off produces a more comparable number of sub-Neptunes and super-Earths, resulting in a better match to the observed relative occurrence rates.

The difference between the two radius distributions decreases with increasing metallicity, as higher-metallicity envelopes are less susceptible to boil-off. At $100\times$ solar, the difference in the number of super-Earths reduces to 20\%. This apparent small invariance, however, results from a compensation between the two effects: a significant fraction of planets still undergo boil-off, but the mass loss is weaker, while models without boil-off retain larger radii and consequently experience stronger photoevaporation at young ages. At $300\times$ solar, boil-off becomes uncommon and is relevant primarily for very low-mass ($\lesssim2M_\oplus$) or highly irradiated ($\gtrsim300F_\oplus$) planets.

\begin{figure}
\centering
\includegraphics[width=0.5\textwidth]{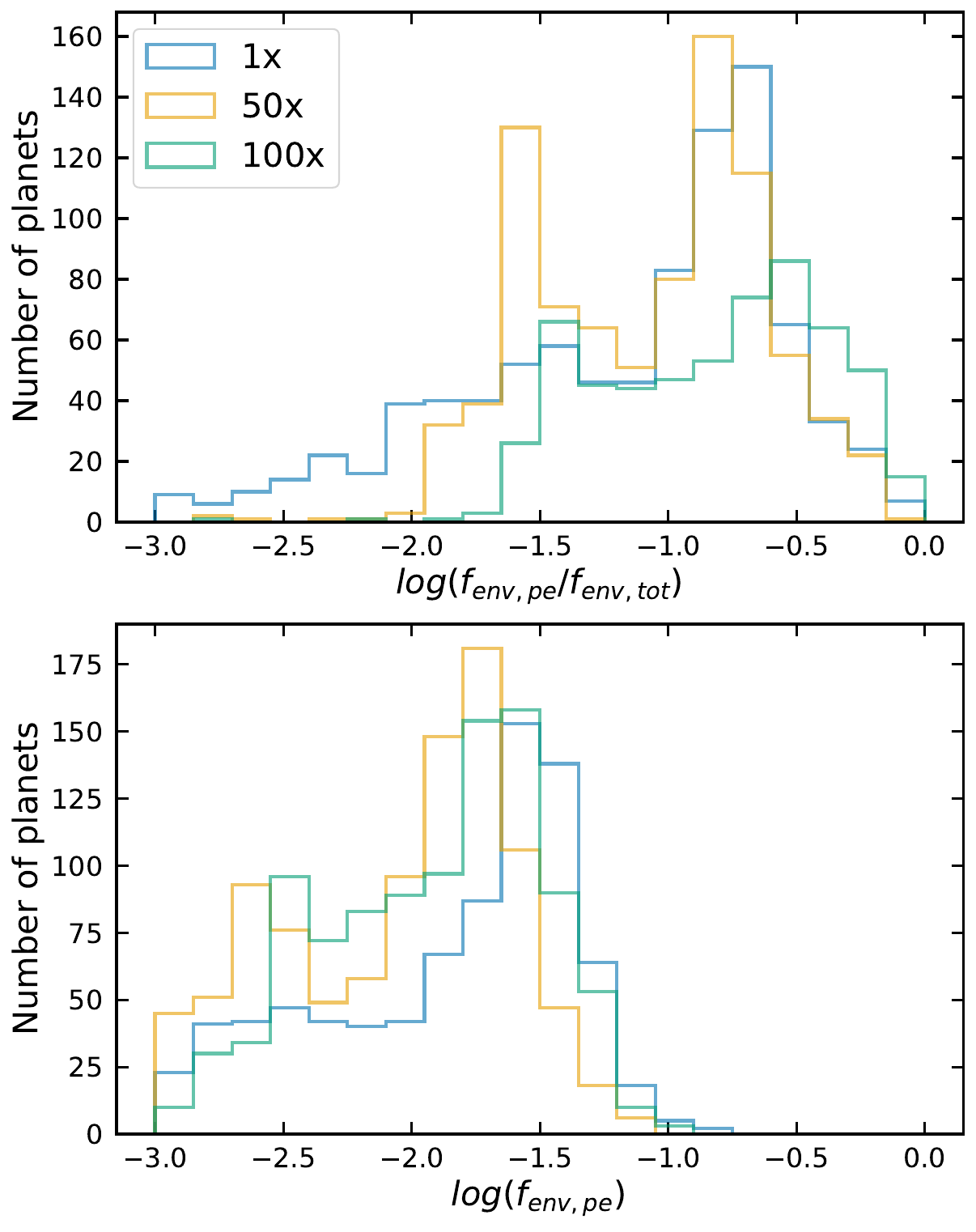} 
\caption{ Relative contribution of photoevaporation to the total mass loss (top) and absolute mass loss due to photoevaporation, normalized by the total planetary mass (bottom). Remarkably, the distributions in the bottom panels consistently peak at around $3\mathrm{wt}\%$. We show results for a range of envelope metallicities and adopt $\zeta=1$.
}
\label{fig:percent}
\end{figure}

In Figure \ref{fig:percent}, we evaluate the relative contributions of photoevaporation and boil-off to the total mass loss for models with the default initial envelope mass ($\zeta=1$). The top panel shows the fraction of the total envelope mass loss attributable to photoevaporation, $f_{\rm env,pe}$, relative to the combined mass loss from photoevaporation and boil-off, $f_{\rm env,tot}$. Here, we include only planets that undergo a boil-off phase. The bottom panel shows the absolute value of $f_{\rm env,pe}$ for the full population. Both panels are shown on logarithmic scales, and we include results for a range of envelope metallicities.

We find that photoevaporation typically accounts for only 10-30\% of the total mass loss among planets that undergo boil-off (top), indicating that boil-off is generally the dominant mass-loss mechanism. For planets that do not undergo boil-off, typically because they are more massive or less irradiated, photoevaporation remains weak despite being the dominant escape mechanism. As shown in the bottom panel, photoevaporation typically removes $\lesssim$3wt\% envelope mass, which is insufficient to completely strip these planets by itself. For comparison, the initial envelope mass fraction is $\sim$10wt\% on average in our fiducial model and $\sim$20wt\% for the default initial envelope mass predicted by the formation model. We therefore suggest that atmospheric mass loss from sub-Neptunes is primarily driven by boil-off, while photoevaporation may commonly complete the final transformation into super-Earths at later ages. Boil-off alone may not fully explain the observed radius distribution, as the final $\sim1$wt\% of envelope mass can critically affect the planetary radius near the radius gap. This result is consistent with the predictions of \citet{Tang24,Tang25b}.

We find that the $\sim$3wt\% limit on photoevaporative mass loss is insensitive to both envelope metallicity and initial envelope mass fraction. However, the relative contributions of boil-off and photoevaporation shift toward photoevaporation as the initial envelope mass decreases, because planets with smaller initial envelopes are less susceptible to boil-off. Overall, boil-off dominates the mass loss for initial envelope mass fractions $\gtrsim$6wt\% (or $\zeta\gtrsim0.3$).

\section{Discussion} \label{sec:disc}
\subsection{Core-Powered Mass Loss}
Over the past decade, the term core-powered mass loss has been widely used to describe bolometric-driven Parker-wind escape. The key argument is that the core plays two essential roles in driving the outflow \citep{Ginzburg16,Ginzburg18,Gupta19,Misener21}: first, it sets the mass-loss rate by regulating the rate at which the envelope re-equilibrates; and second, it can enhance the mass-loss rate by delaying envelope contraction. The first argument was critically examined and objected by \citet{Tang24}. In this work, we provide insight into the second argument by directly quantifying the energetics and mass-loss enhancement.

Based on these results, we argue that the term \emph{core-powered mass loss} is misleading for several reasons. First, as demonstrated in Section \ref{subsec:core}, core luminosity supplies only a small fraction ($\sim10$-$20\%$) of the thermal energy available for envelope re-equilibration and therefore produces only a modest enhancement of bolometric-driven mass loss, corresponding to an absolute increase of $\sim0.4\mathrm{wt}\%$ in our fiducial population. Second, envelope re-equilibration and its associated enhancement of mass loss are not unique to the bolometric-driven phase; in fact, we find that core luminosity has a larger effect on photoevaporation than on boil-off. Third, substantial boil-off occurs even in the absence of core luminosity. Fourth, core luminosity begins to produce a significant enhancement of boil-off only when the planet is already close to the transition to photoevaporation, either after losing $\gtrsim90\%$ of its initial envelope mass or when it begins with a radius close to the transition boundary. Finally, we find that the net effect of core luminosity has only a minimal impact on the planetary radius distribution at the population level.

Several additional sources of confusion have also arisen from the original analytical formulation, naming, and interpretation of core-powered mass loss. In particular, the outflow is driven by stellar heating from the star, while core luminosity primarily affects the interior structure and thermal evolution, thereby setting the lower boundary of the outflow. This distinction is obscured by the terminology. Moreover, the original analytical treatments \citep{Ginzburg18,Misener21} make several simplifying assumptions that can affect the inferred importance of core luminosity, including the omission of the radiative atmosphere, assumptions about the mass-loss rate, and an energy-conservation treatment that can overestimate the amount of core energy retained by the envelope. Detailed discussions are provided in \citet{Tang24} and Section \ref{subsec:atmo}.

We therefore suggest that \emph{core-powered mass loss} should not be used to denote a distinct mass-loss mechanism. Rather, core luminosity is a physical effect that can modify the mass-loss rate of an outflow. The terminology for a mass-loss mechanism should instead reflect either its dominant energy source or its characteristic evolutionary timescale. We therefore prefer to use \emph{bolometric-driven escape} to distinguish this process from photoevaporation, and \emph{boil-off} to distinguish its short-lived phase from long-lived mass loss.

\subsection{Volatile Dissolution} \label{subsec:volatile}
Volatile dissolution and outgassing \citep{Chachan18,kite19,Schlichting22,Tang26} are important physical processes that can affect the sub-Neptune population distribution. There are three key implications for interpreting our population results in light of the potential role of volatile outgassing.

First, for a primordial envelope, hydrogen may dominate the volatile inventory dissolved in the molten rock/iron interior because of its high abundance and strong chemical affinity for the molten planetary interior. Recent simulation work \citep{Stixrude25} suggests that hydrogen can become fully miscible with the magma ocean beyond a critical boundary. This substantial ingassing can shift the radius cutoff toward smaller radii \citep{kite19}. \citet{Tang26} further show that in the fully miscible regime, the remaining atmosphere becomes helium-dominated, producing a radius cutoff near $4R_\oplus$, consistent with the observations \citep{Fulton18,Dattilo23}. If volatile dissolution and outgassing were included in our population models, this effect could allow a larger initial envelope mass, potentially shifting our preferred $\zeta$ from 0.5 toward 1, in better agreement with planet formation models.

Second, \citet{Tang26} show that volatile outgassing can shift the sub-Neptune peak toward smaller radii while preserving the relative abundance of sub-Neptunes and super-Earths at intermediate-to-low solubilities. If this effect is included in our models, it would therefore tend to favor a lower envelope metallicity than our fiducial value of $100\times$ solar. Notably, the $50\times$ solar model in Figure \ref{fig:met} may provide a closer match to the observations in terms of the relative occurrence of sub-Neptunes and super-Earths.

Third, the radius gaps produced by our population models are, on average, deeper than the observed gap. This discrepancy may also be alleviated by volatile outgassing. As shown by \citet{Heng25}, smaller-radius sub-Neptunes tend to have higher envelope metallicities, which can increase the envelope mass retained by these planets relative to our models. This additional envelope mass could partially fill the radius gap and thereby reduce its depth.

In summary, population models that neglect volatile dissolution and outgassing may not fully reproduce the observed population, even if the underlying distributions of ${M_{\rm ric},P,M_s}$ are correctly specified. Within our framework, incorporating volatile dissolution and outgassing introduces trends that can bring the modeled population closer to the observations. A self-consistent treatment of these processes within population synthesis is therefore an important direction for future work.

\section{Conclusions} 
In this work, we present a population synthesis framework based on SEAMIST evolution framework and construct a grid of population models to explore how the planetary population distribution depends on a range of parameters and physical processes. Our main findings are summarized as follows:
\begin{itemize}
    \item Envelope metallicity strongly affects the planetary radius distribution. We find a preferred envelope metallicity of 50-100x solar, which reproduces the observed radius gap and the relative occurrence of sub-Neptunes and super-Earths.
    \item Boil-off generally dominates atmospheric mass loss over photoevaporation. Photoevaporation typically removes $\lesssim3\mathrm{wt}\%$ of the initial envelope mass, and for planets that undergo boil-off, it generally contributes less than $30\%$ of the total mass loss.
    \item Sub-Neptunes exhibit a wide range of mass-loss timescales during photoevaporation, suggesting that the commonly adopted $\sim100$ Myr photoevaporation timescale versus the previously assumed several-Gyr timescale for core-powered mass loss, which we instead find to be $\lesssim$3 Myr, may not reliably distinguish the physical origin of observed outflows.
    \item During bolometric-driven escape (boil-off), core luminosity typically supplies only 10-20\% of the thermal energy available for envelope re-equilibration, corresponding to an absolute mass-loss enhancement of $\sim0.4\mathrm{wt}\%$. The contribution of core luminosity is substantially larger during photoevaporation. Nevertheless, the net effect of core luminosity on long-term planetary radius evolution and the resulting population distribution is small.
    \item We identify a boil-off-photoevaporation transition at $R_{\rm rcb}/R_s\sim$0.05-0.1, where $R_{\rm rcb}$ and $R_s$ are the envelope radius and the isothermal sonic radius, respectively. This transition is largely insensitive to core luminosity, and the long-term radius distribution is also insensitive to the precise transition time, provided that the transition is imposed when the mass-loss timescale is longer than the thermal contraction timescale and at $t\lesssim$5Myr.
    \item Both the radiative atmosphere and boil-off, which are commonly neglected or simplified in previous population studies, can significantly affect the resulting planetary population distribution.
    \item We infer a typical initial envelope mass fraction of $\sim10\%$ pre-boil-off, approximately three times larger than values commonly adopted in previous population studies. This value is more consistent with predictions from planet formation models.
    \item The inferred core-mass distribution is more concentrated toward low-mass cores than the distribution inferred from the observed planetary masses, suggesting that the underlying core-mass distribution may differ substantially from the observed total mass distribution.
\end{itemize}
Given its modest contribution to boil-off, we therefore do not refer to boil-off or bolometric-driven escape as \emph{core-powered mass loss}. In addition, we suggest that core-powered mass loss is better understood as a modification to atmospheric escape through its influence on the lower-boundary conditions, rather than as a distinct mass-loss mechanism.

Even with refined treatments of the radiative atmosphere, boil-off, and envelope metallicity compared to previous simplified models, our best-fit population cannot fully reproduce the observed occurrence-radius distribution, indicating that additional physical processes are needed in population synthesis models. We find that volatile outgassing produces the trend needed to improve agreement with the observed distribution and highlight its importance for inclusion in future studies.

\bibliography{reference}{}
\bibliographystyle{aasjournal}



\end{document}